\documentclass[12pt]{article}
\usepackage{amssymb,amsmath,natbib,graphicx,subcaption,changepage,relsize,amsthm,enumerate,bbm}
\usepackage{placeins}
\usepackage{colortbl}
\usepackage{tikz,url}
\usepackage[margin=1in]{geometry}

\theoremstyle{definition}

\newcommand{\Prob}{\mathbb{P}}

\newcommand{\bmu}{\boldsymbol\mu}
\newcommand{\bnu}{\boldsymbol\nu}

\newcommand{\by}{\boldsymbol{y}}
\newcommand{\bbeta}{\boldsymbol{\beta}}

\newcommand{\beps}{\boldsymbol\epsilon}
\newcommand{\balpha}{\boldsymbol\alpha}

\begin{document}

\begin{center}
{\Large\bf Network Autocorrelation Models with Egocentric Data}

\vspace{0.5in}

Daniel K. Sewell
\footnote{Daniel K. Sewell is Assistant Professor, Department of Biostatistics,
University of Iowa, Iowa City, IA 52242 (E-mail: {\it daniel-sewell@uiowa.edu}).}
\end{center}

\begin{abstract}
Network autocorrelation models have been widely used for decades to model the joint distribution of the attributes of a network's actors.  This class of models can estimate both the effect of individual characteristics as well as the network effect, or social influence, on some actor attribute of interest.  %
Collecting data on the entire network, however, is very often infeasible or impossible if the network boundary is unknown or difficult to define.  Obtaining egocentric network data overcomes these obstacles, but as of yet there has been no clear way to model this type of data and still appropriately capture the network effect on the actor attributes in a way that is compatible with a joint distribution on the full network data.  This paper adapts the class of network autocorrelation models to handle egocentric data.  The proposed methods thus %account for the complex dependencies induced by the entire network, and can estimate the network effect on the actor attribute of interest.  
incorporate the complex dependence structure of the data induced by the network rather than simply using ad hoc measures of the egos' networks to model the mean structure, and can estimate the network effect on the actor attribute of interest.  The vast quantities of unknown information about the network can be succinctly represented in such a way that only depends on the number of alters in the egocentric network data and not on the total number of actors in the network.  Estimation is done within a Bayesian framework.  A simulation study is performed to evaluate the estimation performance, and an egocentric data set is analyzed where the aim is to determine if there is a network effect on environmental mastery, an important aspect of psychological well-being.

%For egocentric networks, however, when the variable of interest is on the actor level, rather than the dyadic level, there has been no clear way to appropriately incorporate network effects.  This paper proposes a novel way of dealing with egocentric data where the variable of interest is on the actor level rather than the dyadic level.  The proposed approach is rooted in the network autocorrelation models, and so utilizes the joint distribution of all actors of the network.  Estimation is done within a Bayesian framework.
\vspace{ 2mm}

\end{abstract}

\section{Introduction}
Network autocorrelation models can help capture complex dependencies in individual level data and can also estimate how and to what extent an individual's network influences that individual's attributes or behaviors.  \cite{fujimoto2011network} describes the network autocorrelation model as ``a workhorse for modeling network influences on individual behavior.''  \cite{wang2014statistical} states ``The network autocorrelation model has some clear advantages over other conventional approaches (e.g., egocentric or dyadic) in that it simultaneously accommodates network effects and individual attributes.''  This class of models has been used for decades in a variety of contexts, such as determining the network effect on gender roles in labor \citep{white1981sexual}, educational and occupational aspirations \citep{duke1991peer}, U.S. interstate commodity flows \citep{chun2012modeling}, policy influence \citep{carpenter1998strength}, task performance \citep{carr2015network}, and phylogenetics \citep{bjorklund1990phylogenetic}.

Network autocorrelation models describe stochastic data generating processes using the joint distribution of all actor attributes given the network structure.  This is both a benefit and a curse.  The positive aspect of this, and indeed the motivation for employing such an approach, is that by jointly modeling all actors in the network, the complex dependence structure is explicitly modeled, and social influence can be directly quantified and estimated.  The downside is that to utilize such a model, one needs to collect data on all actors in the network.  This can be a problem for (at least) three reasons.  First, often times the network is simply too large to sample \citep{granovetter1976network}, or there are monetary constraints to obtaining data on all the actors of the network.  Second, the actors of the network may not be easily accessible to the researchers, especially if the network is defined by controversial or illegal behaviors.  Third, the boundary of the network may not be identifiable.  For example, suppose one wishes to know the network effect of peers on adolescent behaviors.  Is the network of interest defined by all adolescents in a particular class or school?  Or perhaps it can be defined by some on-line social media platform?  Or perhaps it is all adolescents in a particular city, state, or country?  \cite{doreian1989network} makes the statement, which still holds true today, ``locating [a network's] boundaries remains a persistent and vexing problem.''

Researchers often avoid the difficulty of collecting data on all actors of the network by obtaining a subsample of the actors and focusing on the ties involving the sampled actors.  The resulting data is referred to as egocentric network data.  This type of data can be collected in a variety of ways, such as a simple random sample, targeted sampling, snowball sampling, respondent-driven sampling, etc \citep[see, e.g.,][]{heckathorn1997respondent}.  Egocentric network analyses have been used to study interorganizational collaborations \citep{ahuja2000collaboration}, health behaviors \citep{o2012egocentric}, personal and group communication \citep{fisher2005using}, contraceptive use \citep{behrman2003Social}, support network after cancer diagnoses \citep{ashida2009changes}, and many others.

The use of egocentric data has been limited primarily to the study of either dyadic relationships or structural/positional measures of the entire network \citep{provan2007interorganizational}.  Methods to study actor attributes using egocentric data are more limited in scope; this type of analysis is often done in an ad hoc manner by using as a covariate some summary statistic of the egos' personal networks such as density, network size, or an average of some alter attribute.  

This paper proposes a novel method that adapts the network autocorrelation model to egocentric network data.  The proposed method is derived directly from the joint distribution of all actors in the network, even if the boundary of the network is unknown or ill-defined, and thus incorporates the complex dependence structure of the data induced by the network rather than simply using ad hoc measures of the network in the mean structure.  Estimation is done within a Bayesian framework.  
% and performs satisfactorily, though suffering from the same problems that plague network autocorrelation models in general.  
%We hope that this is a first step toward providing a framework for incorporating network effects into a model aimed at egocentric network data which takes into consideration the joint distribution of the entire network.

Section \ref{methods} describes the proposed methodology.  Section \ref{simulationStudy} describes a simulation study that compares the performance of the proposed method with OLS estimators which ignore the network effect and with estimators using the entire network data.  Section \ref{appliedExample} shows the results from applying the proposed method to an egocentric data set of adults in a rural southeastern Iowa town, with the goal of determining if there is a network effect on environmental mastery.  Section \ref{discussion} provides a brief discussion.

\section{Methods}
\label{methods}
Suppose that we wish to make inference regarding a graph augmented with actor attributes.  We may view this as a triple ${\cal G}=({\cal V},{\cal E},{\cal A})$;  ${\cal V}$ is the set of vertices, or actors, of the network, ${\cal E}$ is the set of edges, or relations, between the vertices, and ${\cal A}$ is the set of actor attributes on ${\cal V}$.  We will denote $|{\cal V}|$, the number of actors, by $n$.  Typically one may represent ${\cal E}$ by an adjacency matrix $A$, where the $i^{th}$ row $j^{th}$ column entry of $A$ is 1 if there is an edge between actors $i$ and $j$ and 0 otherwise.  The actor attributes ${\cal A}$ can be partitioned into the $n\times1$ response variable vector $\by$ and the $n\times p$ matrix of covariates $X$.  The goal is to try to determine how the covariates $X$ and the network affect the response $\by$.  This is typically accomplished via network autocorrelation models.

The network autocorrelation model has its genesis in spatial statistics \citep[e.g.,][]{ord1975estimation,doreian1980linear}.  It was soon borrowed by researchers studying complex network data to great effect \citep[e.g.,][]{dow1982network}.  There are two variations on a theme, namely, \citep[borrowing nomenclature from][]{doreian1980linear}, the network effects model, given by
\begin{align}
\by=&X\bbeta + \rho A\by + \beps,&
\label{netEffects}
\end{align}
and the network disturbances model, given by
\begin{align}\nonumber
\by&=X\bbeta +\bnu,&\\
\bnu&=\rho A\bnu + \beps,&
\label{netDisturbances0}
\end{align}
where $\bbeta$ is the parameter vector of coefficients, $\rho$ is the coefficient which captures the network effect, and $\beps$ is a vector of zero mean independent normal random variables with homogeneous variance $\sigma^2$.  Note that an equivalent but more concise form of (\ref{netDisturbances0}) is
\begin{align}
%(I-\rho A)\by&=(I-\rho A)X\bbeta + \beps.&
\by&=X\bbeta +\rho A(\by-X\bbeta)+\beps.&
\label{netDisturbances}
\end{align}

\begin{figure}[t]%[h]%[p]%
\centering
\includegraphics[width=0.5\textwidth]{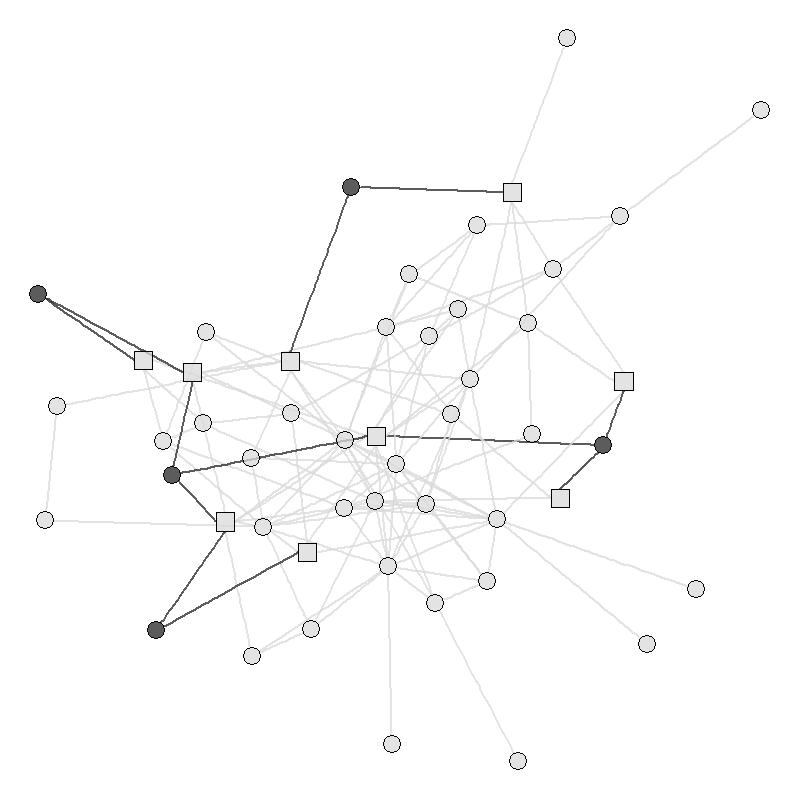}
\caption{A toy example of an egocentric network, where the dark circles are the sampled egos, the dark edges are the observed egos' edges, the squares are the alters, and the light gray circles and lines are the unobserved actors and edges respectively.}
\label{exampleEgo}
\end{figure}

For egocentric data, ${\cal G}$ is only partially observed.  Figure \ref{exampleEgo} illustrates an egocentric network for a small toy data set.  The set of actors ${\cal V}$ can be partitioned into the sampled egos ${\cal V}_e$, the egos' alters ${\cal V}_a$ (those actors with whom the egos have ties), and all other actors in the network ${\cal V}_o$, so that ${\cal V}={\cal V}_e\cup{\cal V}_a\cup{\cal V}_o$.  Let $n_e$, $n_a$, and $n_o$ denote the number of egos, the number of the egos' alters, and the number of remaining actors in the network respectively, so that $n=n_e+n_a+n_o$.  %Note that $n_o$, and hence $n$, is assumed finite but unknown.  
We can partition ${\cal E}$ by focusing on the adjacency matrix $A$, specifically 
\begin{align*}
A&=\begin{pmatrix}
A_e&A_{ea}&{\bf 0}\\
A_{ea}'&A_a&A_{ao}\\
{\bf 0}&A_{ao}'&A_o
\end{pmatrix},
\end{align*}
and similarly we can partition $\by$ and $X$ by
\begin{align}\nonumber
\by& = (\by_e',\by_a',\by_o')'&\\ \nonumber
X&=(X_e',X_a',X_o')',&
\end{align}
where the subscripts $e$, $a$, and $o$ correspond to the egos, the egos' alters, and all other actors in the network respectively.

Obviously there are quite a few unknowns here, not the least of which is the number of others $n_o$, and hence the size of the network $n$.  Specifically, we do not know $\by_a$, $\by_o$, $X_o$, $A_a$, $A_{ao}$, nor $A_o$.  Trying to directly employ either the network effects or disturbances model is clearly not possible with so much missing data (and quite possibly an unknown amount of missing data).  The goal, then, is to capture as much of the information as possible while confining all the unknowns in as few terms as possible.  A conditional distribution, rather than the full joint, will then be used, treating these unknown terms as nuisance parameters to be estimated.  We first show how to do this for the network effects model, and then in a similar manner show the same for the network disturbances model.

\subsection{Network effects model}
We first rewrite (\ref{netEffects}) into the following three separate equations:
\begin{align}
\by_e&=X_e\bbeta+\rho A_e\by_e + \rho A_{ea}\by_a + \beps_e,&\label{netEffects1} \\
\by_a&=X_a\bbeta + \rho A_{ea}'\by_e + \rho A_a\by_a + \rho A_{ao}\by_o +\beps_a&\label{netEffects2}\\
\by_o&=X_o\bbeta + \rho A_{ao}'\by_a + \rho A_{o}\by_o +\beps_o\label{netEffects3}
\end{align}
Note that (\ref{netEffects3}) includes no information about the observed data, and so is not used in deriving the conditional likelihood of the observed data.  From (\ref{netEffects1}) and (\ref{netEffects2}) we obtain
\begin{align*}
\by_e&=(\rho A_e + \rho^2 A_{ea}A_{ea}')\by_e + (X_e+\rho A_{ea}X_a)\bbeta&\\ &+\rho^2A_{ea}\big( A_a\by_a+A_{ao}\by_o \big) + \rho A_{ea}\beps_a+ \beps_e,
\end{align*}
which implies that 
\begin{align}\nonumber
\by_e|\balpha&\sim N\Big(
M_1^{-1}\big[
(X_e+\rho A_{ea}X_a)\bbeta + \rho^2A_{ea}\balpha
\big],\sigma^2M_1^{-1}M_2M_1^{-1}
\Big),& \\ \nonumber
M_1& = I-\rho A_e - \rho^2 A_{ea}A_{ea}'&\\
M_2& = I+\rho^2A_{ea}A_{ea}'&
\label{condDistNetEffects} 
\end{align}
where $\balpha=A_a\by_a+A_{ao}\by_o$.  The unknown (nuisance) parameter $\balpha$ can be viewed as the social influence on the alters that cannot be attributed to the egos; we will refer to $\balpha$ as the residual influence effects.  This $n_a\times1$ unknown parameter vector succinctly sums up all unknown information about the entire network that pertains to $\by_e$, and does not depend on $n_o$.

\subsection{Network disturbances model}
For the network disturbances model, we can derive a conditional distribution similar to (\ref{condDistNetEffects}).  As before, we first rewrite (\ref{netDisturbances}) as three separate equations:
%\begin{align}
%\by_e&=\rho A_e\by_e + \rho A_{ea}( \by_a -X_a\bbeta)+ (I-\rho A_e)X_e \bbeta + \beps_e&\label{netDist1} \\
%\by_a&=\rho A_{ea}'\by_e + \rho A_{a}\by_a +\rho A_{ao}\by_o+ \big(X_a - \rho A_{ea}'X_e -\rho A_{a}X_a - \rho A_{ao} X_o\big)\bbeta + \beps_a&\label{netDist2} \\
%\by_o&=\rho A_{ao}'\by_a + \rho A_{o} \by_o + \big(X_o - \rho A_{ao}'X_a -\rho A_{o}X_o\big)\bbeta + \beps_o\label{netDist3} &\\
%\end{align}
\begin{align}
\by_e&=X_e \bbeta +\rho A_e(\by_e-X_e\bbeta) + \rho A_{ea}( \by_a -X_a\bbeta)+ \beps_e&\label{netDist1} \\
\by_a&=X_a\bbeta + \rho A_{ea}'(\by_e-X_e\bbeta) + \rho A_{a}(\by_a-X_a\bbeta) +\rho A_{ao}(\by_o-X_o\bbeta)+ \beps_a&\label{netDist2} \\
\by_o&=X_o\bbeta + \rho A_{ao}'(\by_a-X_a\bbeta) + \rho A_{o} (\by_o-X_o\bbeta) + \beps_o\label{netDist3} &
\end{align}
As before, (\ref{netDist3}) is irrelevant to the observed egocentric network data and will be disregarded.  Combining (\ref{netDist1}) and (\ref{netDist2}) yields
\begin{align}\nonumber
(I-\rho A_e-\rho^2A_{ea}A_{ea}')\by_e&=(I-\rho A_e-\rho^2A_{ea}A_{ea}')X_e\bbeta &\\ &+\rho^2A_{ea}\big( A_a(\by_a-X_a\bbeta)+A_{ao}(\by_o-X_o\bbeta) \big) +\rho A_{ea}\beps_a + \beps_e,&
\end{align}
which implies that 
\begin{align}
\by_e|\balpha&\sim N\Big(
X_e\bbeta + 
\rho^2M_1^{-1}A_{ea}\balpha,
\sigma^2M_1^{-1}M_2M_1^{-1}
\Big),&
\label{condDistNetDisturbances}
\end{align}
where $\balpha=A_a(\by_a-X_a\bbeta)+A_{ao}(\by_o-X_o\bbeta)=A_a\bnu_a +A_{ao}\bnu_o$ and $M_1$ and $M_2$ are as given previously.  Similar to the network effects model, the residual influence effects $\balpha$ can be interpreted as the influence on the residuals of the alters that cannot be attributed to the egos.

\subsection{Row Normalization}
The choice of $A$ is not always obvious; a notable paper discussing this topic is \cite{leenders2002modeling}.  We will limit our discussion to a commonly used transformation of the adjacency matrix, namely row normalization (column normalization can be addressed in nearly exactly the same way).  Normalizing the rows such that they each sum to 1 is a practice that has, by some authors, been recommended, and has a long history of implementation \citep[e.g.,][]{ord1975estimation,anselin1988spatial}.  

When the adjacency matrix $A$ has been row normalized, we may think of this as replacing $A$ in (\ref{netEffects}) and (\ref{netDisturbances}) with
\begin{equation}
\begin{pmatrix}
D_e & {\bf 0} & {\bf 0}\\
{\bf 0} & D_a & {\bf 0} \\
{\bf 0}&{\bf 0}&D_o
\end{pmatrix}A,
\end{equation}
where $D_e$ is the diagonal matrix whose $i^{th}$ diagonal entry equals the inverse of $\sum_{j=1}^nA_{ij}$ (the degree of actor $i$), and similarly for $D_a$ and $D_o$.  From an egocentric network, we only know $D_e$.  $D_o$ is absorbed entirely into residual influence effects $\balpha$, and thus does not complicate matters.  $D_a$, however, must be estimated as this term does not disappear.  Specifically, (\ref{condDistNetEffects}) becomes
\begin{align}\nonumber
\by_e|\balpha&\sim N\Big(
\widetilde M_1^{-1}\big[
(X_e+\rho D_eA_{ea}X_a)\bbeta + \rho^2D_eA_{ea}\balpha
\big],\sigma^2\widetilde M_1^{-1}\widetilde M_2(\widetilde M_1')^{-1}
\Big),& \\ \nonumber
\widetilde M_1& = I-\rho D_eA_e - \rho^2 D_eA_{ea}D_aA_{ea}'&\\
\widetilde M_2& = I+\rho^2D_eA_{ea}A_{ea}'D_e&
\label{condDistNetEffectsDa}
\end{align}
and similarly, (\ref{condDistNetDisturbances}) becomes
\begin{align}
\by_e|\balpha&\sim N\Big(
X_e\bbeta + 
\rho^2\widetilde M_1^{-1}D_eA_{ea}\balpha,
\sigma^2\widetilde M_1^{-1}\widetilde M_2(\widetilde M_1')^{-1}
\Big).&
\label{condDistNetDisturbancesDa}
\end{align}

%Column normalization may be done in the same manner. In our simulation study and our applied example, however, we only consider row normalization, as this seems to be the more commonly used transformation of the adjacency matrix $A$.

\subsection{Estimation}
%Using the conditional distributions given in (\ref{condDistNetEffects}) and (\ref{condDistNetDisturbances}) reduces the number of unknowns (excluding model parameters) from $\big(n_a(n_a+1)+n_o(n_o+1)\big)/2 + pn_o + n_an_o$ to $n_a$.  
Using the conditional distributions given in (\ref{condDistNetEffects}) and (\ref{condDistNetDisturbances}) reduces the number of unknowns from $n_a+n_o+pn_o+n_a(n_a-1)/2+n_an_o+n_o(n_o-1)/2$ associated with $\by_a$, $\by_o$, $X_o$, $A_a$, $A_{ao}$, and $A_o$, to just $n_a$ unknowns associated with the residual influence effects $\balpha$.  
While this is a dramatic reduction (especially if $n_o$ is in the thousands or millions), the parameter space of this conditional model is still very high dimensional.  For our simulations and applied example, we performed estimation within a Bayesian framework with some success.  Specifically we implemented a Metropolis-Hastings-within-Gibbs sampler to obtain draws from the posterior.  This is done by first setting the following priors:
\begin{align}
\sigma^2&\sim IG\left( \frac{a}{2},\frac{b}{2} \right),&\\
\bbeta&\sim N({\bf c},D),&\\
\balpha&\sim N({\bf e},F),&\\
\rho&\sim N(g,h),&
\end{align}
where $IG(a/2,b/2)$ is an inverse gamma distribution with shape $a/2$ and scale $b/2$, and $N({\bf a},B)$ is the normal distribution with mean vector ${\bf a}$ and covariance matrix $B$.  For each of the unknown parameters, samples are drawn from the full conditional distributions which, for all except $\rho$, are well known distributions which are conjugate to the priors.  See the appendix for the full conditional distributions.  For $\rho$, we performed a Metropolis-Hastings step using a normal random walk proposal.

Knowing that $\balpha=A_a\by_a + A_{ao}\by_o$ (or $\balpha=A_a(\by_a-X_a\bbeta)+A_{ao}(\by_o-X_o\bbeta)$ for the disturbances model), one may try to construct a more informative prior on $\balpha$, though the exact distribution cannot be known.  For our simulation study and our applied example, however, we kept the prior on $\balpha$ centered at zero with spherical covariance matrix with a large variance component, thus making the prior flat.

When we wish to row normalize $A$, we must also estimate $D_a$.  There are, of course, a variety of ways in which to do this.  In our simulation study and applied example, we put an upper bound on the range of $1/D_a[j,j]$, the degree of alter $j$, equal to the maximum degree observed in the egos plus some constant.  The lower bound is automatically fixed by the $j^{th}$ column sum of $A_{ea}$.  We then assumed a uniform prior on the integers in this range.

An important aspect of estimation is the negative bias on $\rho$.  When the full data is collected, this is still a well known problem with fitting a network autocorrelation model, and has been discussed in, e.g., \cite{dow1982network}, \cite{smith20044spatial}, \cite{mizruchi2008effect}, and \cite{fujimoto2011network}, among others.  This problem is fully present in the context of using a subset of the full network data.  As this is still an unresolved problem in the full data setting, we leave this issue in the context of egocentric network data as an area of future research.

\section{Simulation Study}
\label{simulationStudy}
We simulated 100 networks from a preferential attachment model each with 1,000 actors.  The stochastic model we used adds actors one by one, drawing a degree (number of edges) from a poisson distribution with mean 5 (rescaled such that there is a zero probability of sending zero edges).  Each edge connects with the $i^{th}$ existing actor with probability proportional to $1+deg_i^{0.01}$, where $deg_i$ is the degree of the $i^{th}$ actor.  We fixed $\bbeta=(-1,0,1)$, $\sigma^2=2$, and set $\rho=0.075$ if there was no row normalization, otherwise we set $\rho=0.75$.

We then simulated $\by$ for each network according to (\ref{netEffects}) and also according to (\ref{netDisturbances}), both with and without row normalization.  Thus there were 400 simulated data sets in total.  For each data set, we then obtained a simple random sample of 150 actors and analyzed the egocentric network data using the proposed approach.  To compare, we analyzed the same egocentric data using OLS, ignoring the network effect entirely.  We also compared our results to that obtained from applying either the network effects model or the network disturbances model using the full network data.  This last is hardly a fair comparison, as it uses on an order of magnitude larger number of data points; nevertheless this serves as some baseline as to what optimum performance could be achieved letting $n_e\rightarrow n$ using a Bayesian approach with the same prior distributions.

\begin{figure}[h]%[p]%[h]
\centering
\begin{subfigure}	{0.35\textwidth}
\includegraphics[width=\textwidth]{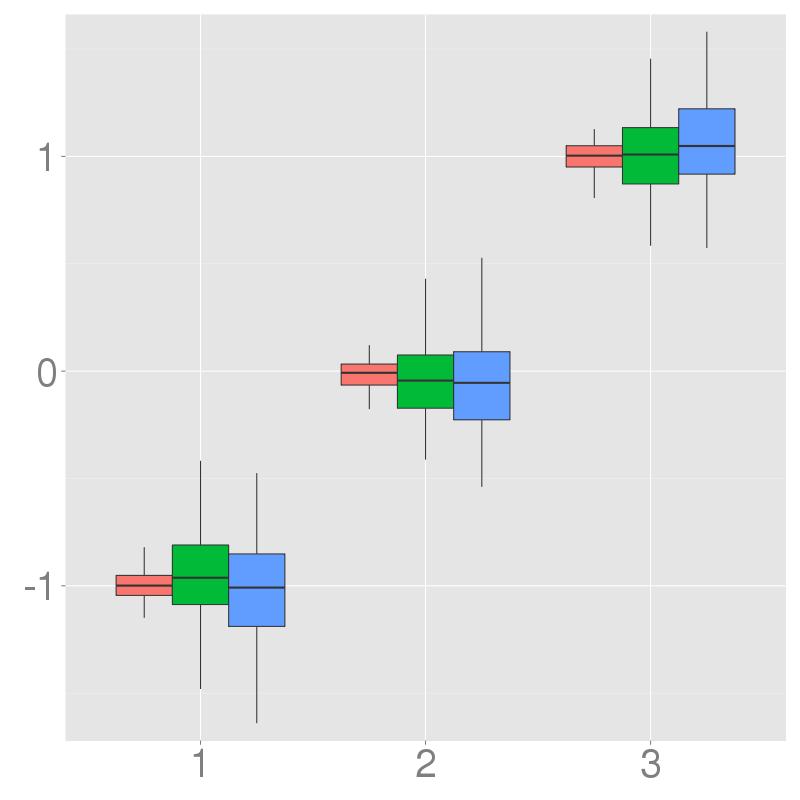}
\caption{Network effects}
\end{subfigure}
\begin{subfigure}	{0.35\textwidth}
\includegraphics[width=\textwidth]{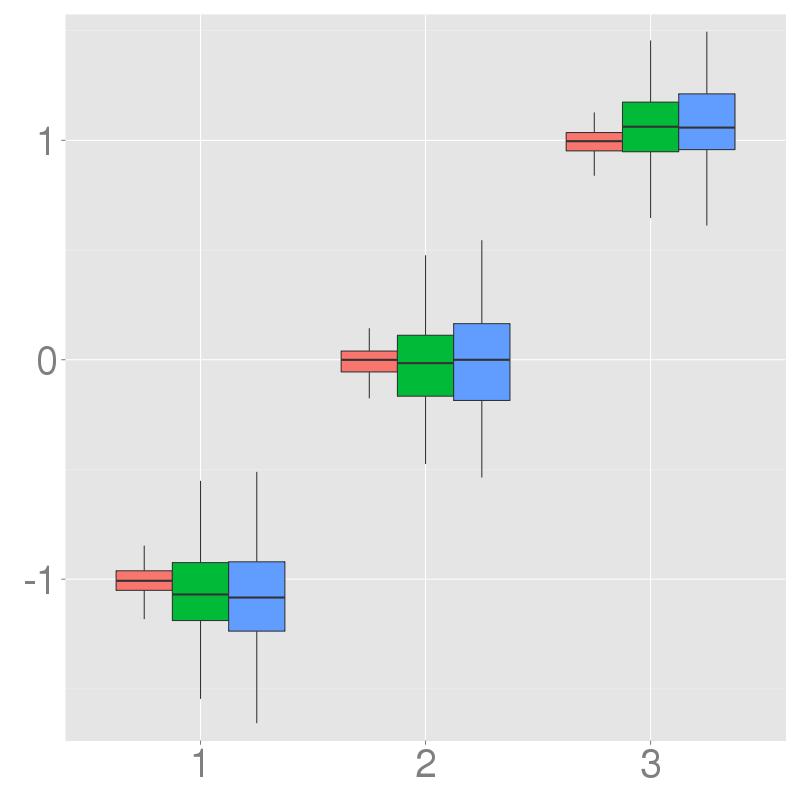}
\caption{Network effects with row normalization}
\end{subfigure}\\
\begin{subfigure}	{0.35\textwidth}
\includegraphics[width=\textwidth]{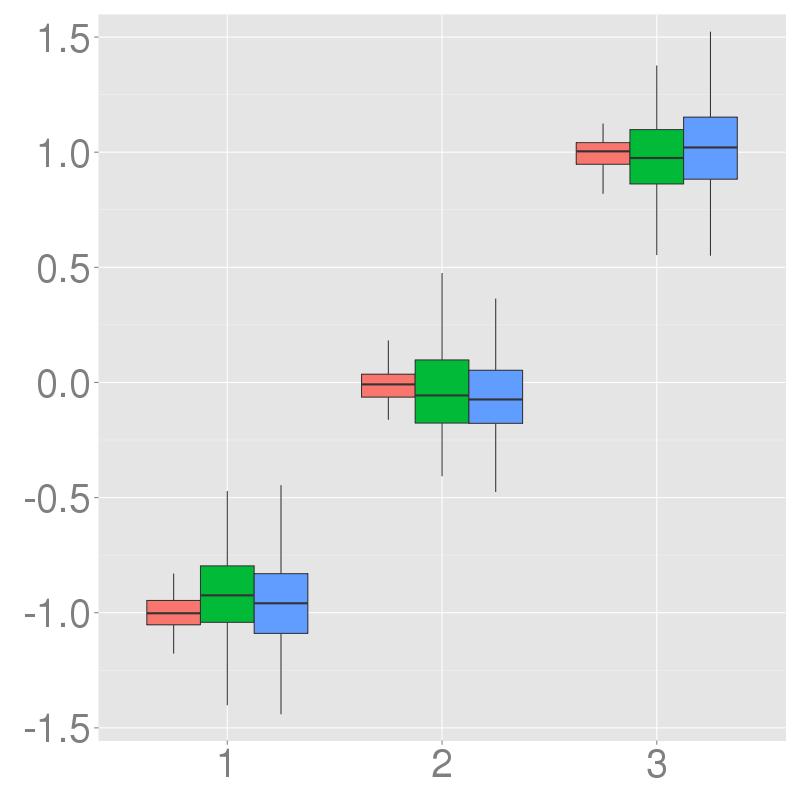}
\caption{Network disturbances}
\end{subfigure}
\begin{subfigure}	{0.35\textwidth}
\includegraphics[width=\textwidth]{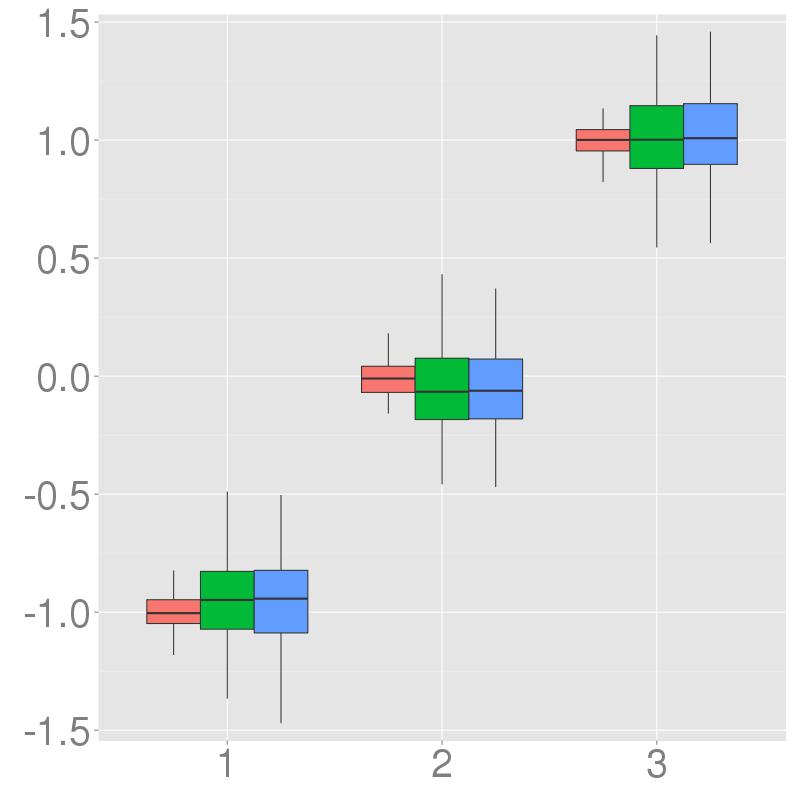}
\caption{Network disturbances with row normalization}
\end{subfigure}
\caption{Simulation study: Estimates for $\bbeta$.  True values are -1, 0 and 1.  For each $\beta_j$, the boxplots correspond to, from left to right, the full data model, the proposed approach, and the OLS estimates ignoring network effects.}
\label{simBetas}
\end{figure}

\begin{figure}[h]%[p]%[h]
\centering
\begin{subfigure}	{0.35\textwidth}
\includegraphics[width=\textwidth]{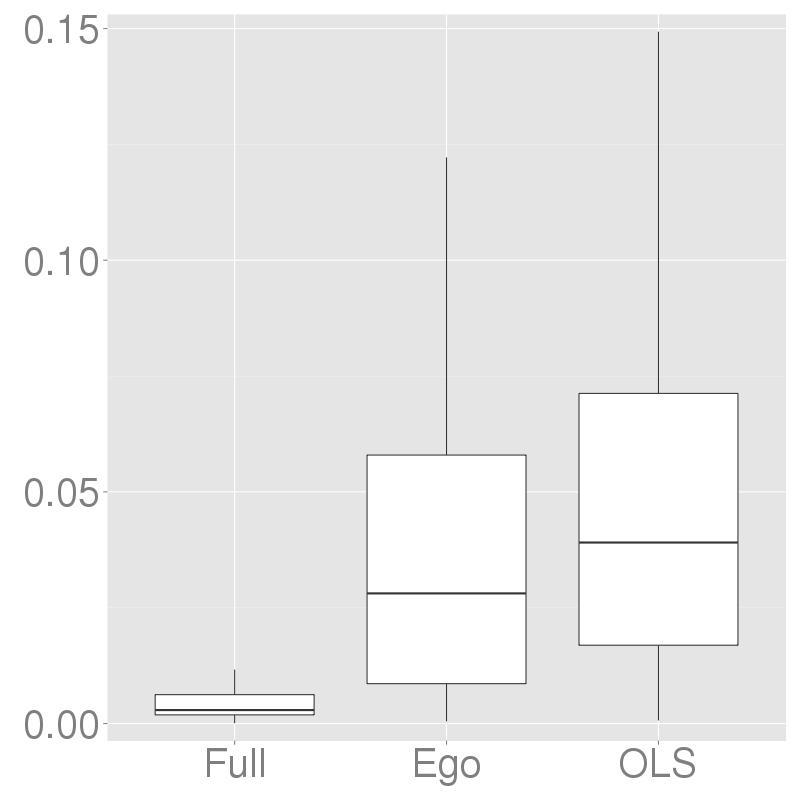}
\caption{Network effects}
\end{subfigure}
\begin{subfigure}	{0.35\textwidth}
\includegraphics[width=\textwidth]{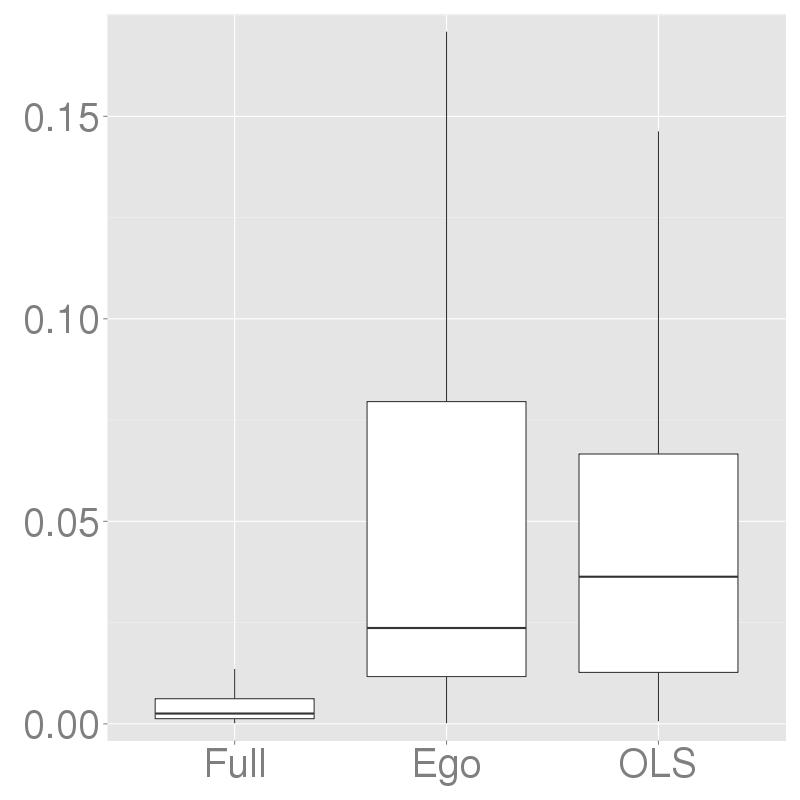}
\caption{Network effects with row normalization}
\end{subfigure}\\
\begin{subfigure}	{0.35\textwidth}
\includegraphics[width=\textwidth]{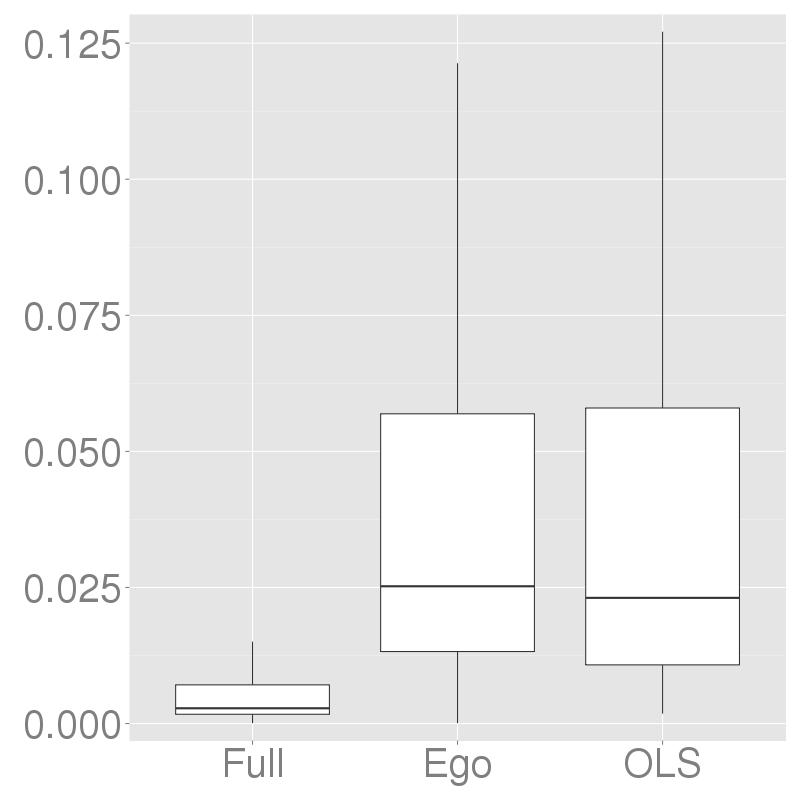}
\caption{Network disturbances}
\end{subfigure}
\begin{subfigure}	{0.35\textwidth}
\includegraphics[width=\textwidth]{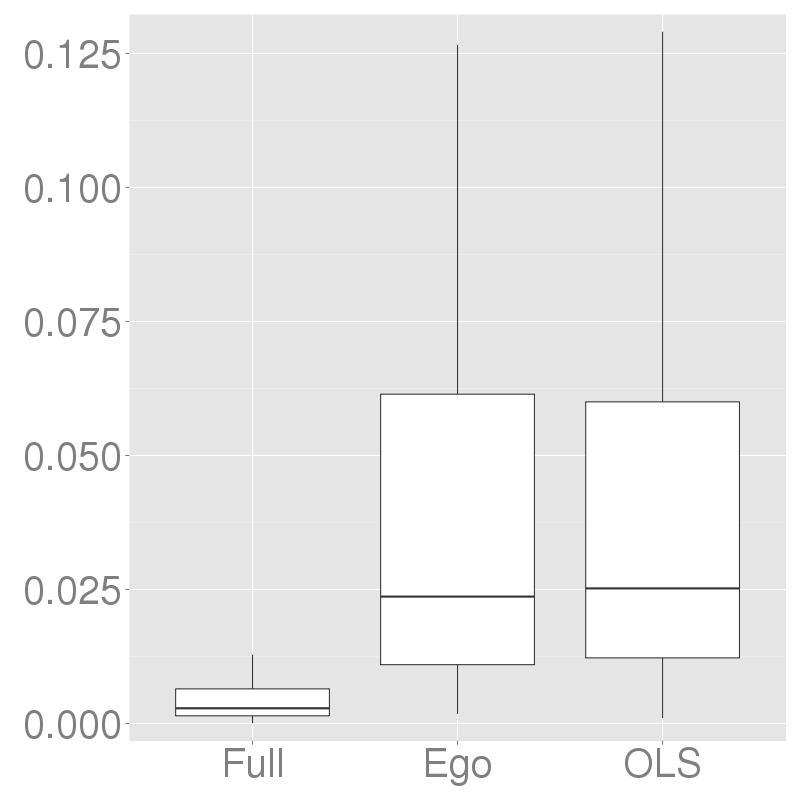}
\caption{Network disturbances with row normalization}
\end{subfigure}
\caption{Simulation study: MSE for $\bbeta$ estimates.}
\label{simMSE}
\end{figure}

\begin{figure}[h]%[p]%[h]
\centering
\begin{subfigure}	{0.35\textwidth}
\includegraphics[width=\textwidth]{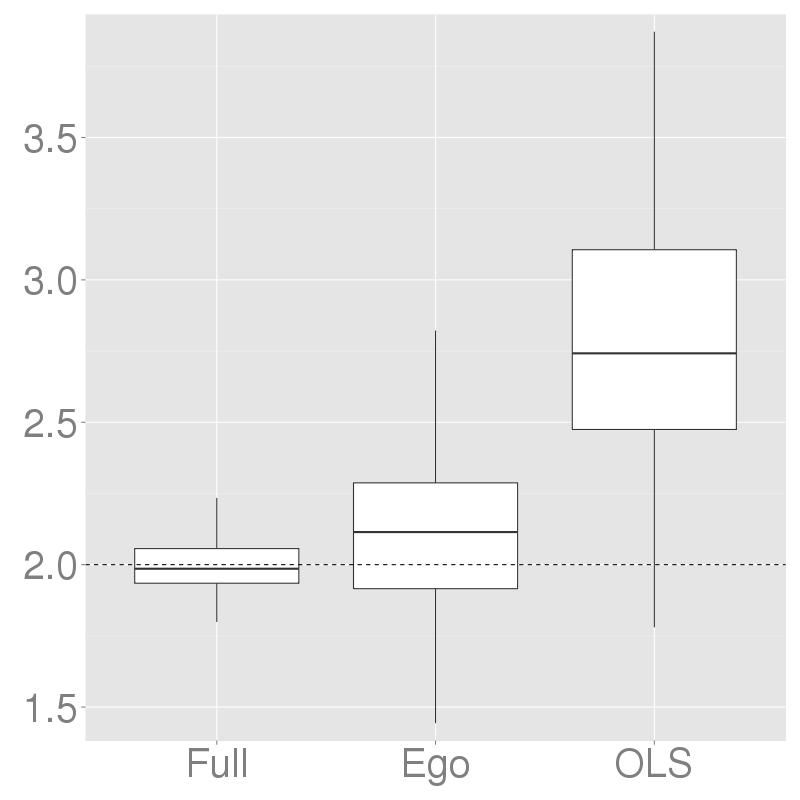}
\caption{Network effects}
\end{subfigure}
\begin{subfigure}	{0.35\textwidth}
\includegraphics[width=\textwidth]{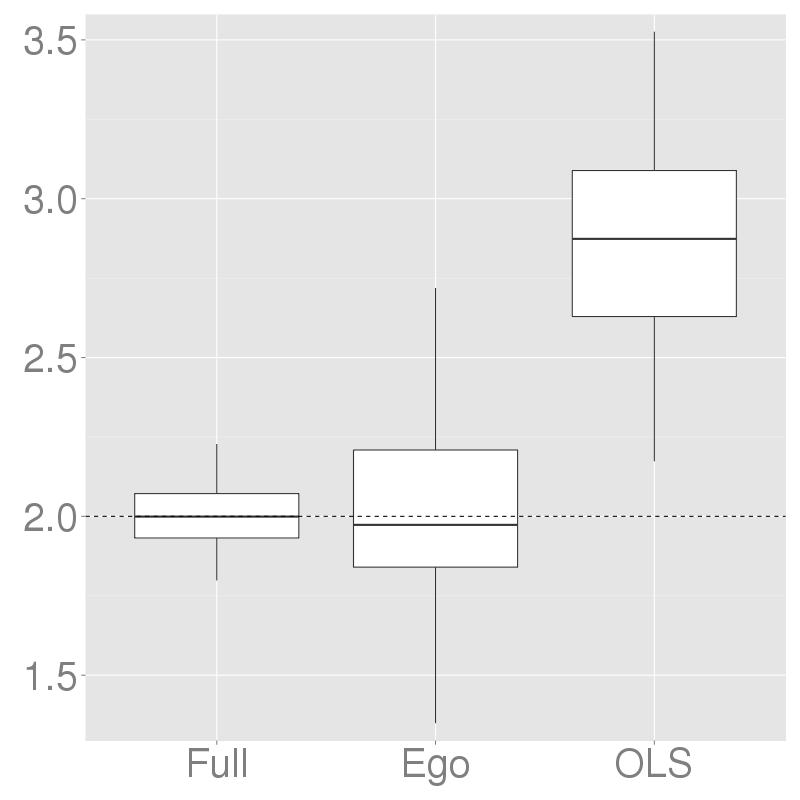}
\caption{Network effects with row normalization}
\end{subfigure}\\
\begin{subfigure}	{0.35\textwidth}
\includegraphics[width=\textwidth]{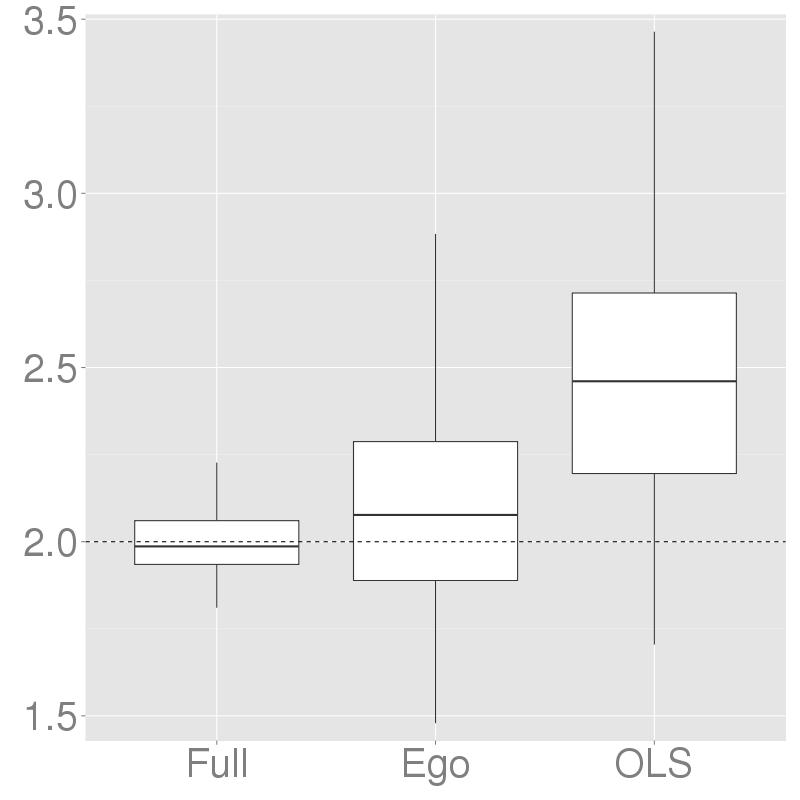}
\caption{Network disturbances}
\end{subfigure}
\begin{subfigure}	{0.35\textwidth}
\includegraphics[width=\textwidth]{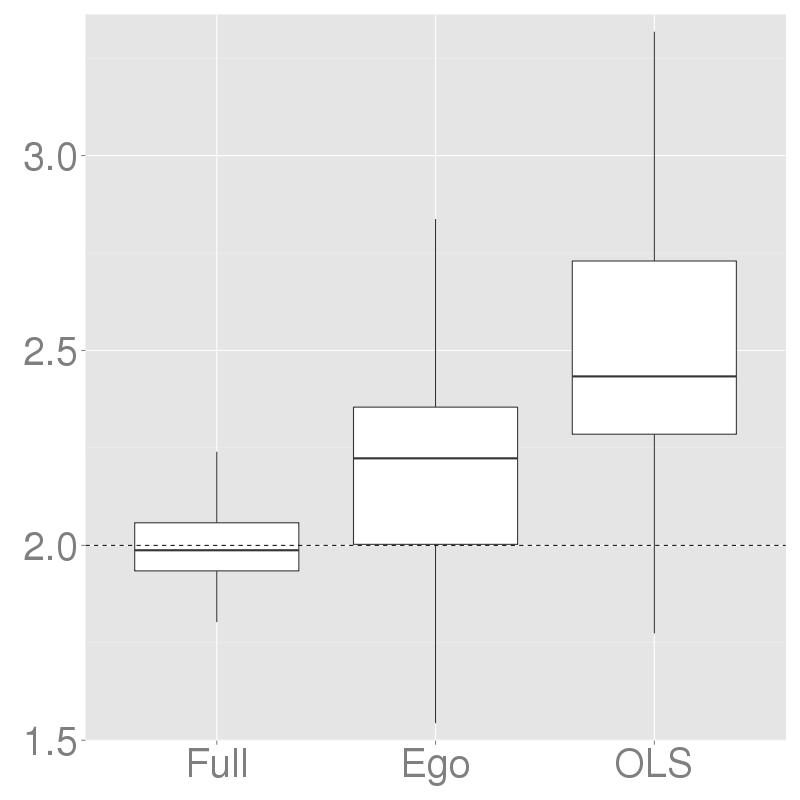}
\caption{Network disturbances with row normalization}
\end{subfigure}
\caption{Simulation study: Estimates for $\sigma^2$.  True value is 2.}
\label{simSig2}
\end{figure}

\begin{figure}[h]%[p]%[h]
\centering
\begin{subfigure}	{0.35\textwidth}
\includegraphics[width=\textwidth]{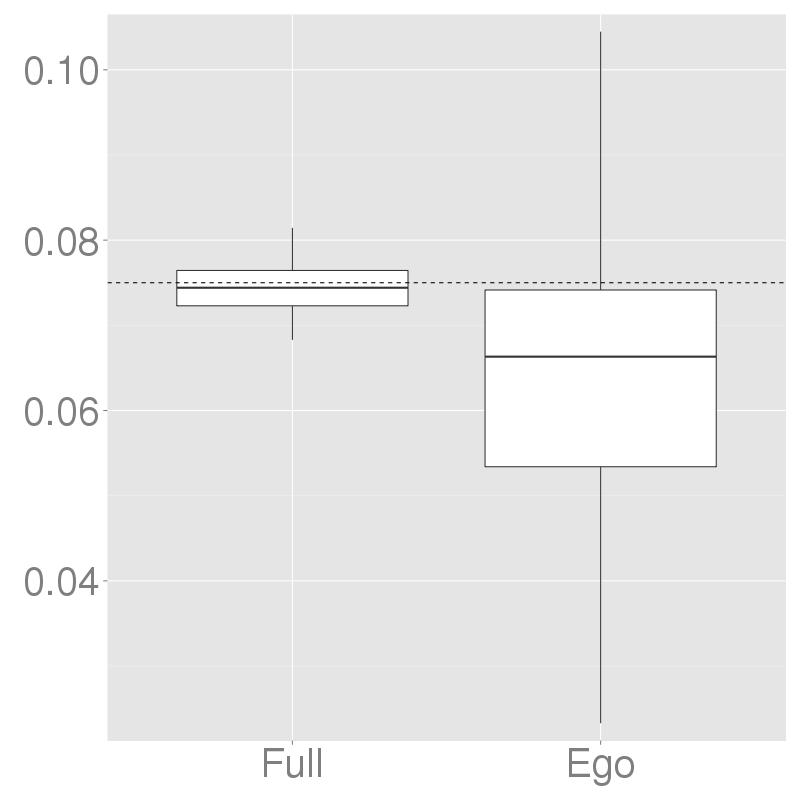}
\caption{Network effects}
\end{subfigure}
\begin{subfigure}	{0.35\textwidth}
\includegraphics[width=\textwidth]{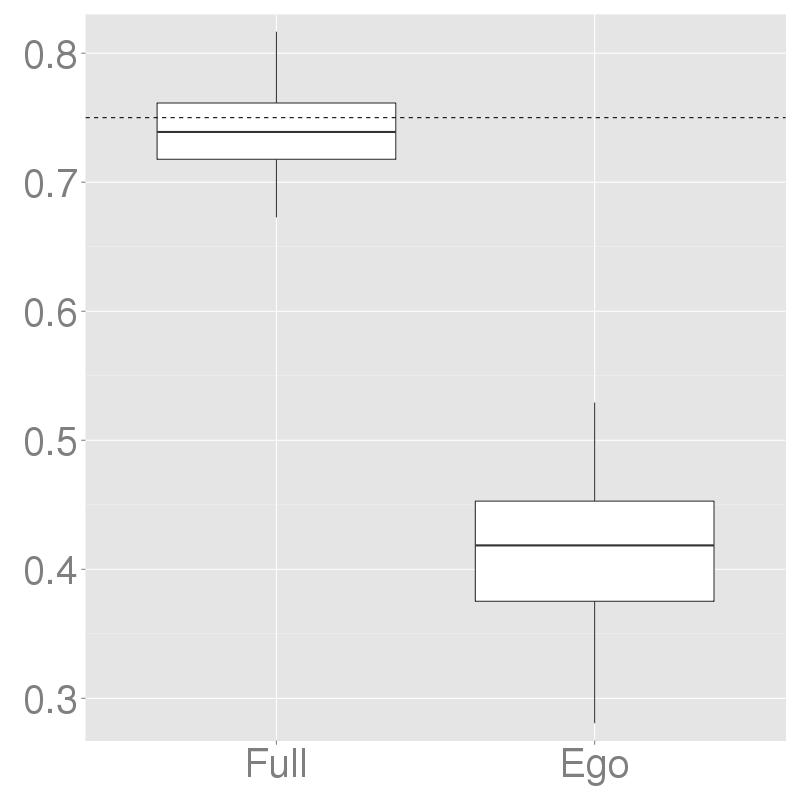}
\caption{Network effects with row normalization}
\end{subfigure}\\
\begin{subfigure}	{0.35\textwidth}
\includegraphics[width=\textwidth]{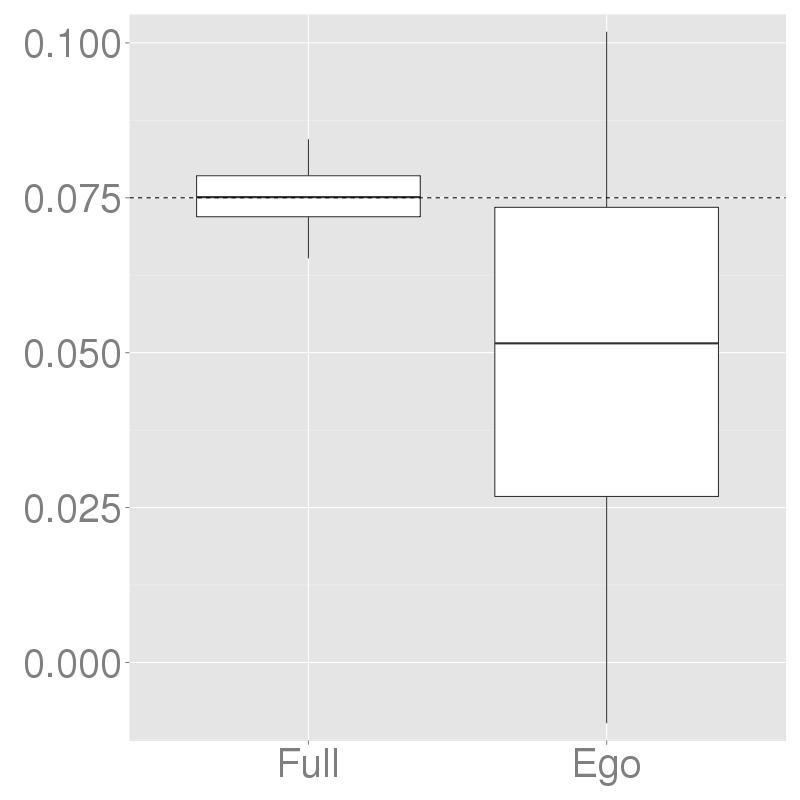}
\caption{Network disturbances}
\end{subfigure}
\begin{subfigure}	{0.35\textwidth}
\includegraphics[width=\textwidth]{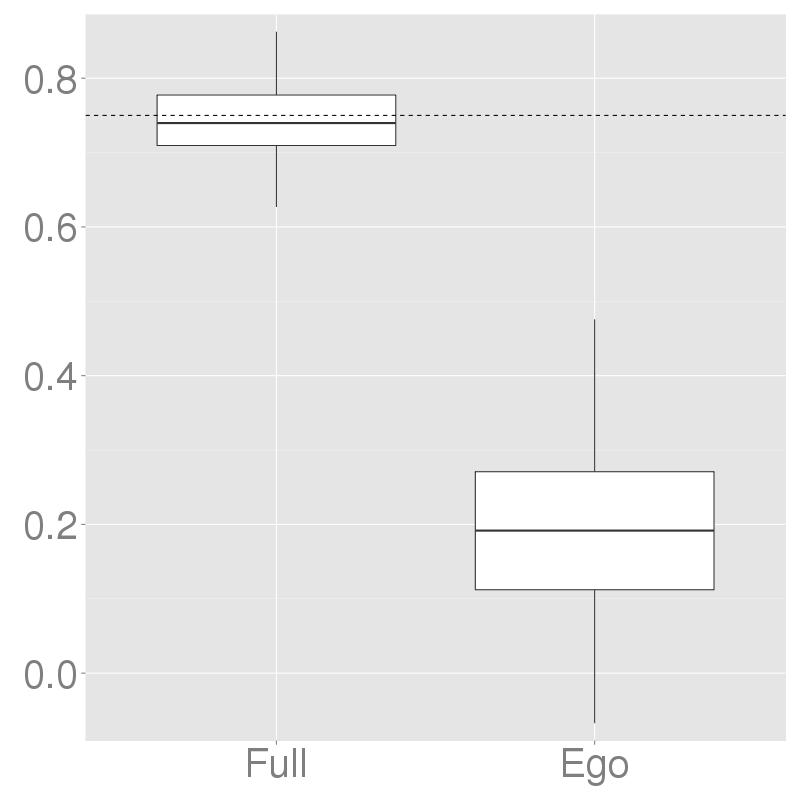}
\caption{Network disturbances with row normalization}
\end{subfigure}
\caption{Simulation study: Estimates for $\rho$.  True value is 0.075, or .75 in the context of row normalization.}
\label{simRho}
\end{figure}

Figures \ref{simBetas} through \ref{simRho} give the results graphically.  Figure \ref{simBetas} gives the boxplots of the point estimates of $\bbeta$ for all simulations, and Figure  \ref{simMSE} shows the boxplots of the MSE, computed for each simulation as $\|\widehat{\bbeta}-\bbeta^{True}\|^2/3$.  From these two figures we see that our proposed method does very similarly to the OLS estimates of the mean coefficients.  Note that for the disturbances model the OLS estimates are unbiased.  Figure \ref{simSig2} shows that ignoring the network effect leads to an upward bias in the OLS variance estimation in all cases.  

From Figure \ref{simRho} we see that our proposed method seems to do a reasonable job of estimating the network effect, although there is, as alluded to earlier and seen in the literature for network autocorrelation models in general, a negative bias.  This bias is much more severe in the models with row normalization.  This is entirely unsurprising, as there is a considerable amount of additional uncertainty due to the unknown alter degrees.  To a much lesser extent, there is also more bias exhibited in the network disturbances model compared with the network effects model.  This too is unsurprising, because the network effects model inherently uses more information than the network disturbances model in estimating the influence not directly attributable to $\by_e$.  That is, in the network effects model, the non-ego influence ($\rho A_{ea}\by_a$ in (\ref{netEffects1})) can in part be explained by the known alter covariate information $X_a$ and observed responses $\by_e$, whereas in the network disturbances model, the non-ego influence ($\rho A_{ea}(\by_a-X_a\bbeta)$ in (\ref{netDist1})) is constructed from unknown residuals ($\bnu$ in (\ref{netDisturbances0})).  

%Even though in all cases the estimate of the network effect $\rho$ experienced negative bias, in the simulated network effects model a 90\% credible region was always strictly contained in $\Re^+$.  In the network disturbances model, this was not the case; only 44\% of the time for the unnormalized and 36\% of the time for the row normalized data was the 90\% credible region strictly contained in $\Re^+$.  Hence we see that there is much less power in detecting network effects using the network disturbances model.  This can again be explained by the fact that the disturbances model is leveraging less information than the effects model.

\section{Environmental mastery in older adults}
\label{appliedExample}
Researchers in the Department of Community and Behavioral Health at the University of Iowa collected a rich egocentric network data set on older adult subjects in a rural southeastern Iowa town \citep[see, e.g.,][]{ashida2016reaching}.  One-time interviews were conducted with individuals in which a large number of attributes were collected as well as a variety of dyadic relationships corresponding to the ego.  Here we make use of a subset of this dataset, focusing on an index that represents an individual's environmental mastery.  Specifically, the questions (given in Table \ref{envMastQs}) are taken from the Ryff scales of psychological well-being \citep{ryff1995structure}.  This is an important aspect of the psychology of older adults, and we wish to investigate the notion that there is a network effect on older adults' environmental mastery after accounting for some basic demographic information.  Specifically, we control for gender, race (white or non-white), and age.  The network under consideration was obtained by looking at, for each ego, all individuals with whom the ego sees at least once a week.

\begin{figure}[h]%[p]%
\centering
\includegraphics[width=0.5\textwidth]{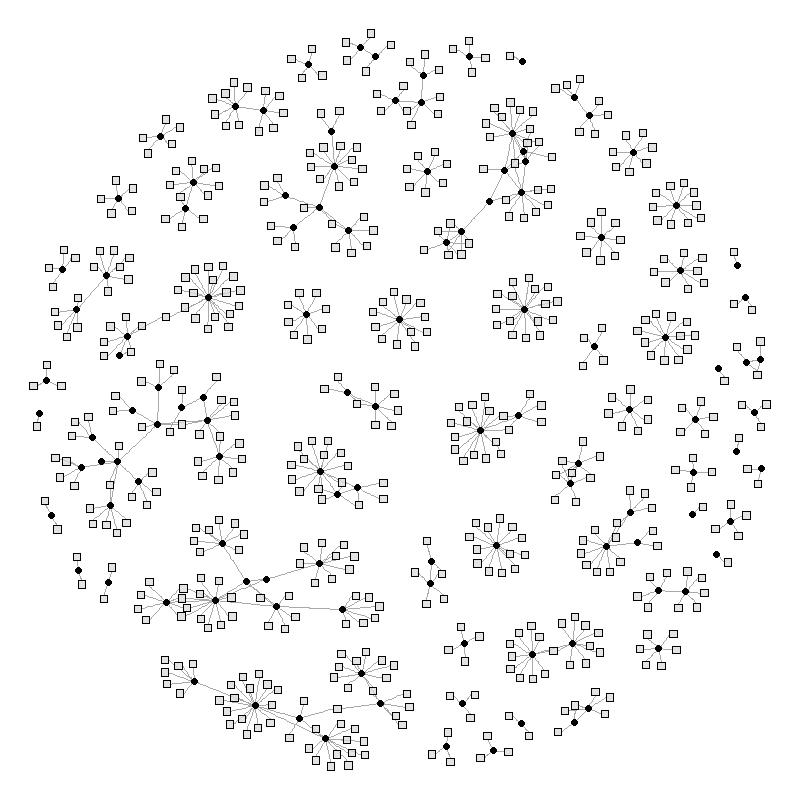}
\caption{Environmental mastery egocentric network.  Black circles are the observed egos.  Gray squares are the alters.}
\label{envMastNetFig}
\end{figure}

There are 119 egos in the data set and a total of 561 alters.  Some of the egos nominated each other, and hence $A_e$ is not a matrix of 0's.  The density of $A_e$ and $A_{ea}$ were both 0.009 after rounding to three decimal places.  The mean degree of the egos was 6.15, ranging from 1 to 18.  Figure \ref{envMastNetFig} shows the network.

We applied both the network effects and disturbances models both with and without row-normalization, running an MCMC algorithm with 300,000 iterations, using 50,000 of these as a burn-in.  Table \ref{envMastLogLikVals} gives the maximum log likelihood values for each of the four implemented models.  Since the network effects model and the network disturbances model have the same number of parameters, and since those models with row normalization have even more parameters (due to estimating the degree distribution of the alters), both AIC and BIC would favor the network effects model without row normalization\footnote{For the four models implemented, there is no nesting, and hence it is perfectly reasonable to find a model with more parameters yielding a lower likelihood than a model with fewer parameters}, and hence further inference is based on this.
%most of the commonly used information criteria would select the network effects model without row normalization\footnote{For the four models implemented, there is no nesting, and hence it is perfectly reasonable to find a model with more parameters yielding a lower likelihood than a model with fewer parameters}.

\begin{figure}[h]%[p]%[h]
\centering
\begin{subfigure}{0.25\textwidth}
\includegraphics[width=\textwidth]{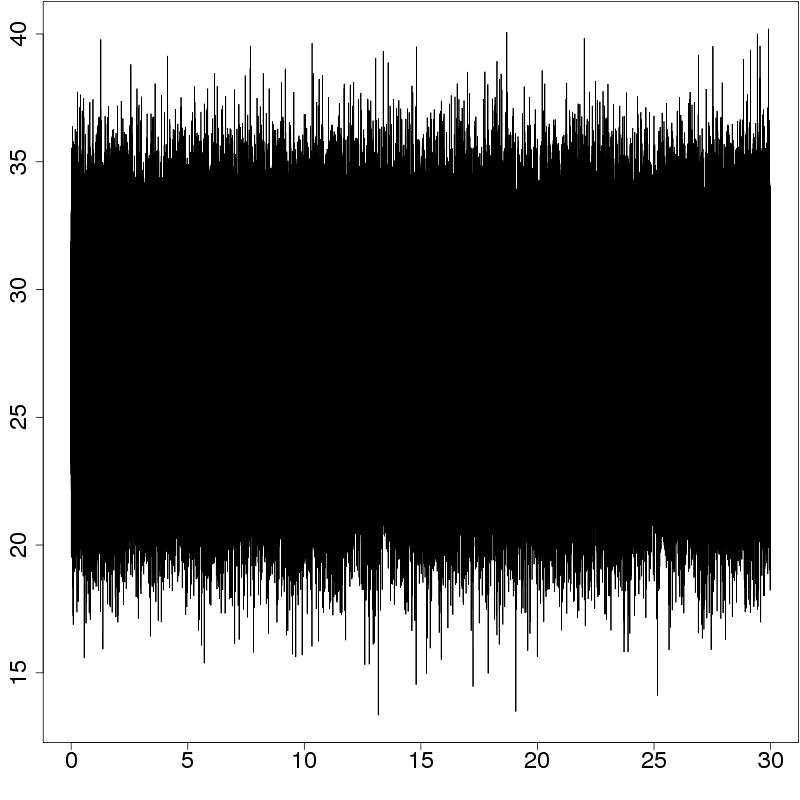}
\caption{Intercept}
\end{subfigure}
\begin{subfigure}{0.25\textwidth}
\includegraphics[width=\textwidth]{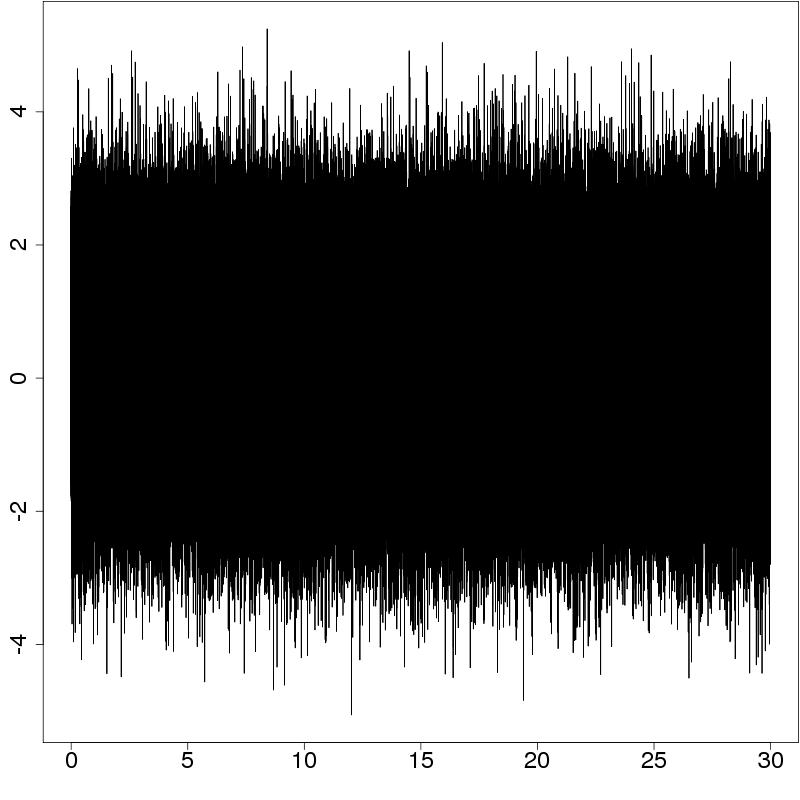}
\caption{Female}
\end{subfigure}
\begin{subfigure}{0.25\textwidth}
\includegraphics[width=\textwidth]{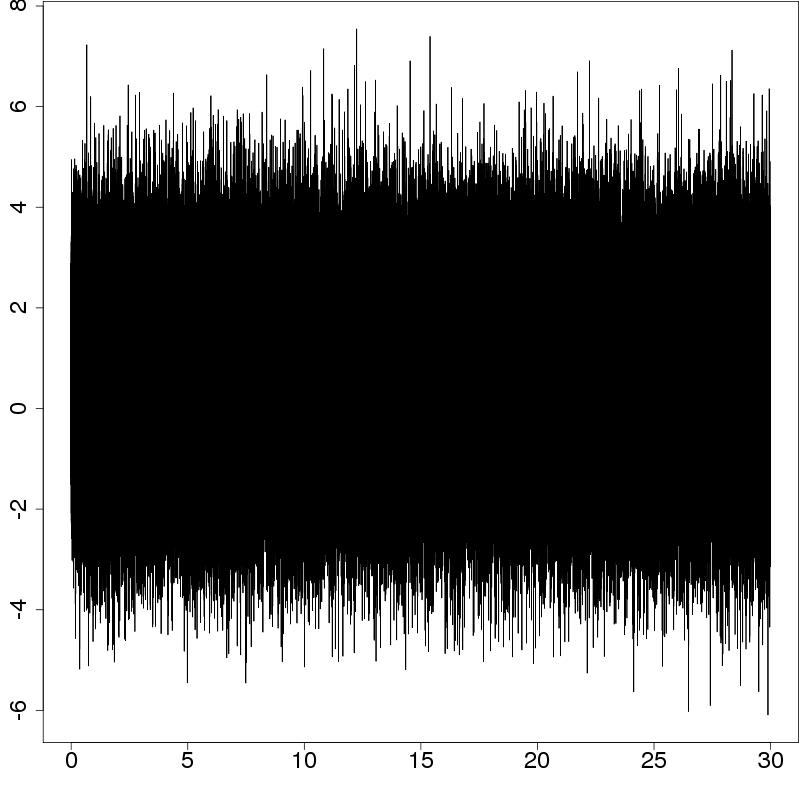}
\caption{Non-white}
\end{subfigure}\\
\begin{subfigure}{0.25\textwidth}
\includegraphics[width=\textwidth]{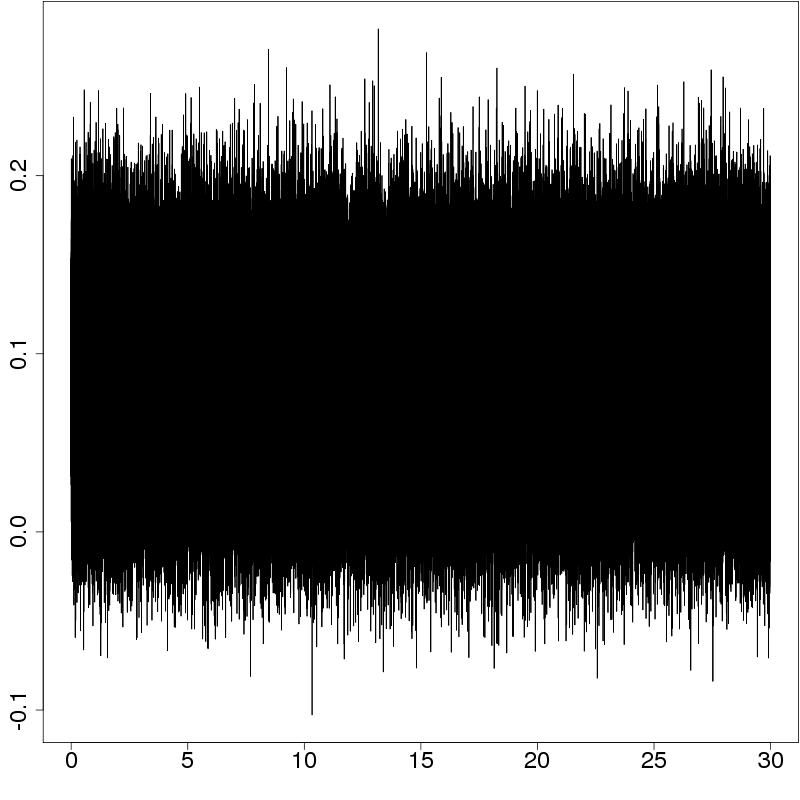}
\caption{Age}
\end{subfigure}
\begin{subfigure}{0.25\textwidth}
\includegraphics[width=\textwidth]{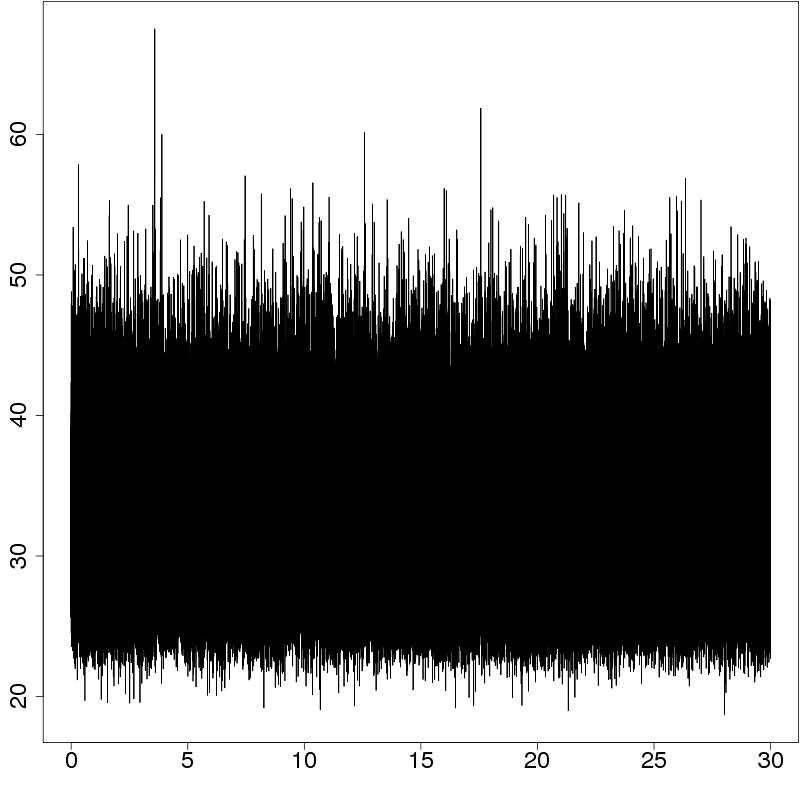}
\caption{$\sigma^2$}
\end{subfigure}
\begin{subfigure}{0.25\textwidth}
\includegraphics[width=\textwidth]{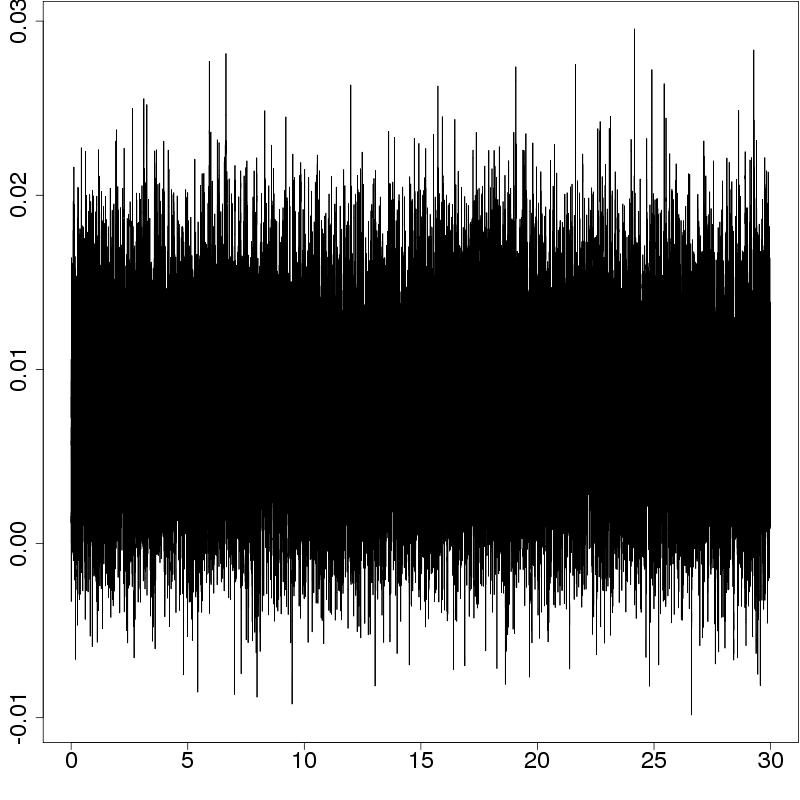}
\caption{$\rho$}
\end{subfigure}
\caption{Environmental mastery data: Trace plots of model parameters.  The x-axis is in 10,000 iterations.}
\label{modARTracePlots}
\end{figure}

\begin{figure}[h]%[p]%
\centering
\begin{subfigure}{0.45\textwidth}
\includegraphics[width=\textwidth]{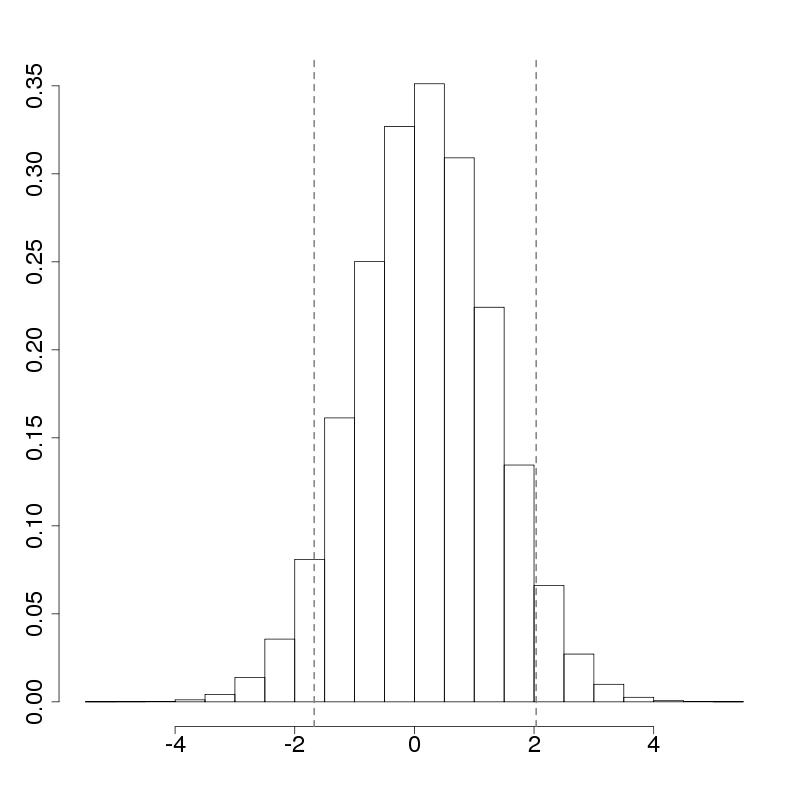}
\caption{Female}
\end{subfigure}
\begin{subfigure}{0.45\textwidth}
\includegraphics[width=\textwidth]{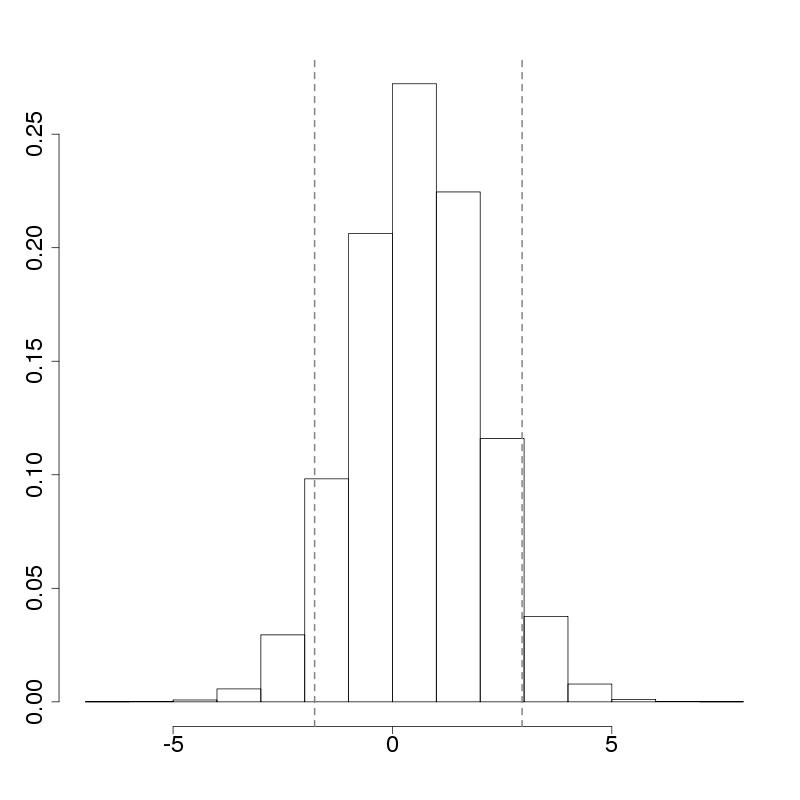}
\caption{Non-white}
\end{subfigure}\\
\begin{subfigure}{0.45\textwidth}
\includegraphics[width=\textwidth]{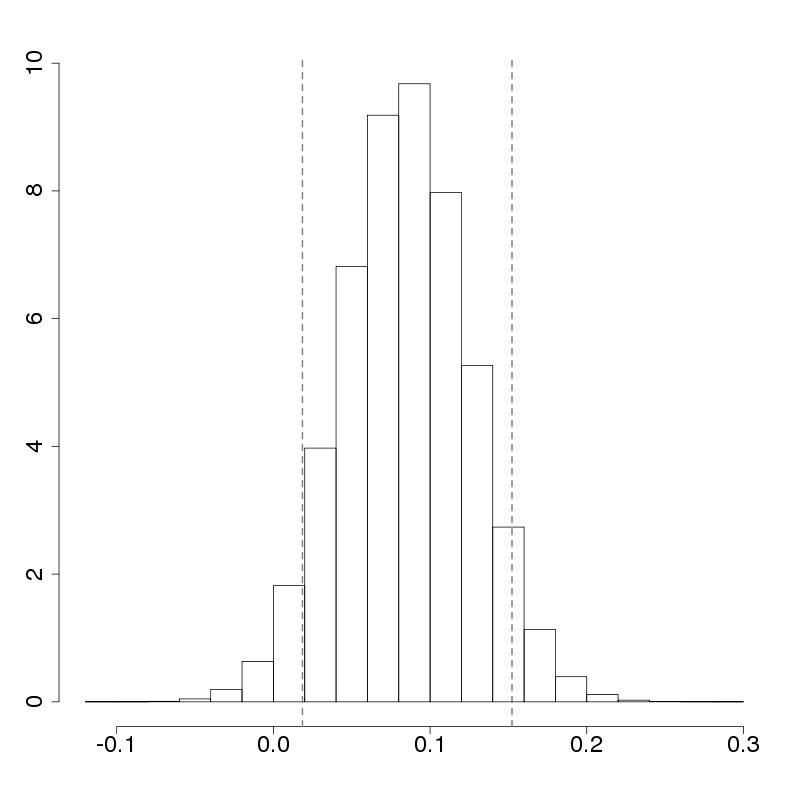}
\caption{Age}
\end{subfigure}
\begin{subfigure}{0.45\textwidth}
\includegraphics[width=\textwidth]{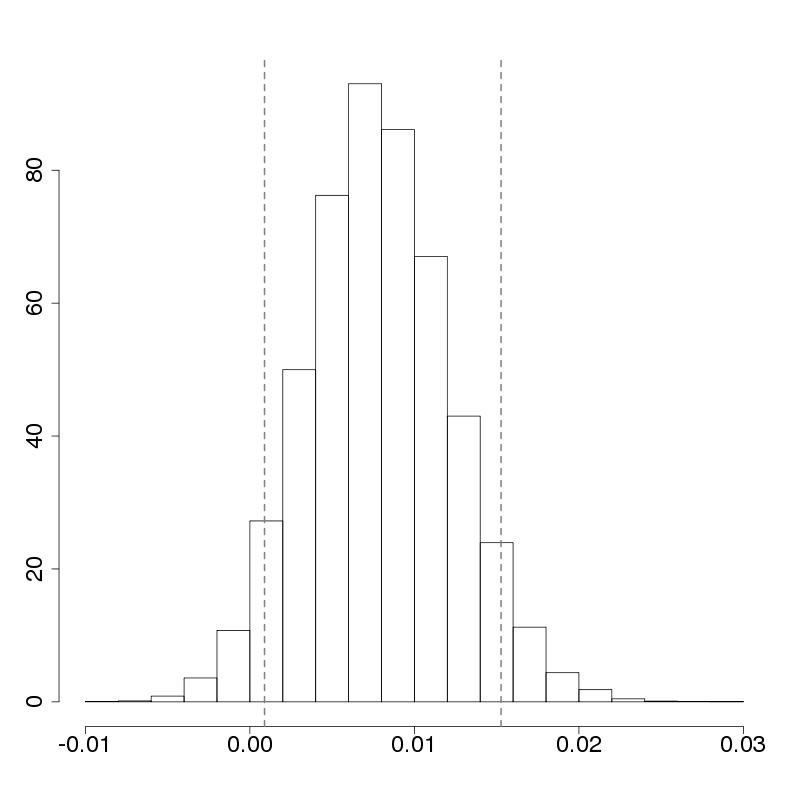}
\caption{$\rho$}
\end{subfigure}
\caption{Environmental mastery data: Histograms of the posterior samples for the mean coefficients for Female, Non-white, and Age, as well as the histogram for $\rho$.  Vertical dotted lines represent the lower and upper bounds for the 90\% credible region.}
\label{modARHists}
\end{figure}

Figure \ref{modARTracePlots} gives the trace plots of the model parameters.  These figures indicate very strongly that the MCMC algorithm converged.  The posterior means and 90\% credible regions for the model parameters are given in Table \ref{envMastPostMeans}, and the histograms of the posterior samples for Female, Non-white, Age, and $\rho$ are given in Figure \ref{modARHists}.  From this we see that of the basic demographics, there is only evidence that age plays a non-trivial role in environmental mastery.  We also see evidence in favor of a network effect.  In fact, there is a 0.969 posterior probability that there is a positive network effect (i.e., $\Prob(\rho>0|\mbox{data})=0.969$).  We may then conclude that an individual's environmental mastery is affected by the environmental mastery with whom the individual is in contact on a regular basis.  
%an individual's environmental mastery is more similar to the environmental mastery of those with whom the individual is in regular contact than would otherwise be expected.  
This type of conclusion may help shape future interventions by focusing on actors with high degree to maximize intervention impact.

\begin{table}
\begin{tabular}{|l|}
\hline
{\bf Environmental Mastery Survey Questions}\\ \hline
1. In general, I feel I am in charge of the situation in which I live.\\ 
2. The demands of everyday life often get me down.\\
3. I do not fit very well with the people and the community around me.\\ 
4. I am quite good at managing the many responsibilities of my daily life.\\ 
5. I often feel overwhelmed by my responsibilities.\\ 
6. I have difficulty arranging my life in a way that is satisfying to me.\\ 
7. I have been able to build a home and a lifestyle for myself that is much to my liking.\\ \hline
\end{tabular}
\caption{List of Likert scale questions used to construct the environmental mastery scores.  Questions 2, 3, 5, and 6 are reverse coded.}
\label{envMastQs}
\end{table}

\begin{table}
\centering
\begin{tabular}{|lr|}\hline
\multicolumn{1}{|c}{{\bf Model}} & \multicolumn{1}{c|}{{\bf Log-likelihood }}\\ \hline
Network effects without row normalization & -30.31\\
Network effects with row normalization & -33.81\\
Network disturbances without row normalization & -31.34\\
Network disturbances with row normalization & -35.65 \\ \hline
\end{tabular}
\caption{Environmental mastery data: Log-likelihood values for each of the four models implemented.}
\label{envMastLogLikVals}
\end{table}

\begin{table}
\centering
\begin{tabular}{|cr|} \hline
{\bf Parameter} &\multicolumn{1}{c|}{{\bf Posterior mean (90\% Credible Interval)}}\\ \hline
Intercept&27.4 (22.6,32.3)\\ \hline
Female&0.178 (-1.68,2.04)\\ \hline
Non-white& 0.589 (-1.77,2.95)\\ \hline
Age & 0.0850 (0.0185,0.152)\\ \hline
$\sigma^2$ & 32.6 (26.2,40.3)\\ \hline
$\rho$ & 0.00788 (0.000894,0.0153)\\ \hline
\end{tabular}
\caption{Environmental mastery data: Posterior means and 90\% credible intervals for the model parameters.}
\label{envMastPostMeans}
\end{table}

%mean(modAR$rho[-burnin]>0)#0.968
%quantile(modAR$rho[-burnin],probs=c(0.025,0.975))#-4.0x10^{-4},0.017
%quantile(modAR$rho[-burnin],probs=c(0.05,0.95))#8.8x10^{-4},0.015
%
%apply(modAR$beta[-burnin,-1],2,quantile,probs=c(0.025,0.975))
%#2.5%  -1.897162 -13.796957 -1.995155 0.002104449
%#97.5%  2.583012   6.100401  3.783459 0.162149063
%apply(modAR$beta[-burnin,-1],2,quantile,probs=c(0.05,0.95))
%#5%  -1.547039 -12.215553 -1.519644 0.01468559
%#95%  2.218915   4.508077  3.308916 0.14891492

\section{Discussion}
\label{discussion}
%Discuss how the net. disturb. model can be used even if you don't have any covariate information on the alters.
Network autocorrelation models are widely used to measure covariate and network effects on a response variable of interest.  These models, however, necessitate data on all actors of the network.  This is very often not feasible.  Egocentric network data are very often dramatically more feasible to collect, but the current methods to estimate network effects on this type of data are ad hoc, and not founded on a data generating process that could explain the full network data and all the complex dependencies therein.  %This paper proposes a method of using network autocorrelation models on egocentric network data.  

This paper derives a model for egocentric data that is consistent with a data generating process that can account for the full network data.  Specifically, if the true underlying generating process is a network autocorrelation model, the proposed conditional distribution used in this paper converges to the joint distribution of the data as $n_e\rightarrow n$.  That is, when $n_e=n$, %we can set all elements of $A$ except $A_e$ to 0 and thus 
%$\by_e=\by$, $A_{e}=A$, $X_e=X$; hence
 (\ref{condDistNetEffects}) is equivalent to the distribution of $\by$ as given in (\ref{netEffects}), and (\ref{condDistNetDisturbances}) is equivalent to the distribution of $\by$ as given in (\ref{netDisturbances}).

The negative bias in the estimation of the network effect as quantified by the parameter $\rho$ is an important issue in network autocorrelation models.  The simulation study has shown that it is present in our context of egocentric data, and is especially problematic when there is row-normalization.  It is the author's hope that this is an area of future research that receives its due attention.

As mentioned earlier, a common ad hoc approach to estimating network effects with egocentric network data is to use as a covariate either network size or an average of some alter attribute.  This can be viewed as using a spatial Durbin model, rather than a more sophisticated network autocorrelation model, only looking at a subset of the data.  The Durbin model for the full data is
$$
\by = X_1\bbeta_1 + AX_2\bbeta_2 + \beps,
$$
where $X_1$ and $X_2$ may share some, all, or none of their columns.  There is no complicated dependence structure in this model (which seems unrealistic in the network context), and so it is straightforward to use this model for egocentric data.  Using the network size as a covariate is equivalent to letting $X_2$ be the vector of 1's.  Using the average of the alter attributes is equivalent to using a row-normalized $A$.  Including lagged exogenous variables can be accounted for in the egocentric network autocorrelation models described in this paper, though some modification is necessary.  Specifically, the network effects model becomes
\begin{align}\nonumber
\by_e|\balpha&\sim N\big(
M_1^{-1}[(X_{1e}+\rho A_{ea}X_{1a})\bbeta_1+((A_e+\rho A_{ea}A_{ea}')X_{2e}+A_{ea}X_{2a})\bbeta_2 +\rho A_{ea}\balpha],&\\
&\hspace{2.5pc}%{30pc}
\sigma^2M_1^{-1}M_2M_1^{-1}\big)&
\end{align}
and similarly the network disturbances model becomes
\begin{align}
\by_e|\balpha&\sim N\big(X_{1e}\bbeta_1 + (A_eX_{2e}+A_{ea}X_{2a})\bbeta_2+\rho^2M_1^{-1}A_{ea}\balpha, \sigma^2M_1^{-1}M_2M_1^{-1}\big).&
\end{align}

%This approach may also be used in the case where the network boundaries are known but there is missing data.  This may be accomplished in a variety of ways that should be context specific, but should always revolve around the appropriate partitioning of $\by$, $X$, and $A$ as given in Section \ref{methods}.

%Recall that the non-ego based influence $\balpha$ in the network disturbances model is based on the residuals $\bnu$, whereas for the network effects model the alter covariates do not play a role in $\balpha$ and instead may be leveraged in parameter estimation.  As explained earlier, this diminishes the ability to determine a network effect in the data.  
Finally, recall that the network effects model leverages more information than the network disturbances model in explaining the non-ego influence, diminishing the ability to determine a network effect in the data.  While this is obviously an important drawback of using the disturbances model rather than the network effects model, there is an important advantage here as well.  Specifically, if it is not possible to measure the covariates on the alters, then while one may not implement the network effects model of (\ref{condDistNetEffects}) (at least not as given here), it is still possible to implement the network disturbances model of (\ref{condDistNetDisturbances}), as $X_a$ has no bearing on the network effect.

\section*{Appendix: Full conditional distributions}
\subsection*{A1: Network effects model}
The full conditional distribution for the variance $\sigma^2$ is given by
\begin{align*}
\sigma^2|\cdot&\sim IG\left(
\frac{a+n_e}{2},\frac12\left[
b + \|M_2^{-\frac12}(M_1\by_e-(X_e+\rho A_{ea}X_a)\bbeta-\rho^2A_{ea}\balpha)\|^2
\right]
\right).&
\end{align*}

\noindent The full conditional distribution for the mean coefficients $\bbeta$ is given by
\begin{align*}
\bbeta|\cdot & \sim N(\bmu_b,\Sigma_b),&\\
\bmu_b&=\Sigma_b\left( 
\frac1{\sigma^2} (X_e + \rho A_{ea}X_a)' M_2^{-1}(M_1\by_e-\rho^2A_{ea}\balpha) +D^{-1}{\bf c}
\right),&\\
\Sigma_b^{-1}&=\frac1{\sigma^2}(X_e + \rho A_{ea}X_a)'M_2^{-1}(X_e + \rho A_{ea}X_a)+D^{-1}.
\end{align*}

\noindent The full conditional distribution for the residual influence effects $\balpha$ is given by
\begin{align*}
\balpha|\cdot &\sim N(\bmu_a,\Sigma_a)&\\
\bmu_a&=\Sigma_a\left(
\frac{\rho^2}{\sigma^2}A_{ea}'M_2^{-1}(M_1\by_e-(X_e+\rho A_{ea}X_a)\bbeta)+F^{-1}{\bf e}
\right), &\\
\Sigma_a^{-1}&=\frac{\rho^4}{\sigma^2}A_{ea}'M_2^{-1}A_{ea} + F^{-1}.&
\end{align*}

\subsection*{A2: Network disturbances model}
The full distribution for the variance $\sigma^2$ is given by
\begin{align*}
\sigma^2|\cdot&\sim IG\left(
\frac{a+n_e}{2},\frac12\left[
b + \|M_2^{-\frac12}(M_1\by_e-M_1X_e\bbeta-\rho^2A_{ea}\balpha)\|^2
\right]
\right).&
\end{align*}

\noindent The full conditional distribution for the mean coefficients $\bbeta$ is given by
\begin{align*}
\bbeta|\cdot & \sim N(\bmu_b,\Sigma_b),&\\
\bmu_b&=\Sigma_b\left( 
\frac1{\sigma^2} X_e'M_1'M_2^{-1}(M_1\by_e-\rho^2A_{ea}\balpha) +D^{-1}{\bf c}
\right),&\\
\Sigma_b^{-1}&=\frac1{\sigma^2}X_e'M_1'M_2^{-1}M_1X_e+D^{-1}.
\end{align*}

\noindent The full conditional distribution for the residual influence effects $\balpha$ is given by
\begin{align*}
\balpha|\cdot &\sim N(\bmu_a,\Sigma_a)&\\
\bmu_a&=\Sigma_a\left(
\frac{\rho^2}{\sigma^2}A_{ea}'M_2^{-1}(M_1\by_e-M_1X_e\bbeta)+F^{-1}{\bf e}
\right), &\\
\Sigma_a^{-1}&=\frac{\rho^4}{\sigma^2}A_{ea}'M_2^{-1}A_{ea} + F^{-1}.&
\end{align*}

\FloatBarrier
\bibliographystyle{harvard}
\bibliography{LNAMForEgocentricData}
\end{document}